\documentstyle[12pt,subeqn,subeqnarray]{article}

\def\be{\begin{equation}}
\def\ee{\end{equation}}
\def\beq{\begin{eqnarray}}
\def\eeq{\end{eqnarray}}
\def\lsim{\:\raisebox{-0.5ex}{$\stackrel{\textstyle<}{\sim}$}\:}
\def\gsim{\:\raisebox{-0.5ex}{$\stackrel{\textstyle>}{\sim}$}\:} 
\def\GeV{{\rm ~GeV}}

\begin{document}
\begin{flushright}
TIFR/TH/03-04 
\end{flushright}
\bigskip
\begin{center}
{\Large{\bf Higgs and SUSY Searches at LHC: An Overview}} \\[3cm]
{\large D.P. Roy} \\[1cm]
Tata Institute of Fundamental Research, \\ Homi Bhabha Road, 
Mumbai 400 005, India
\end{center}
\bigskip\bigskip

I start with a brief summary of Higgs mechanism and supersymmetry.
Then I discuss the theoretical constraints, current limits and search
strategies for Higgs boson(s) at LHC --- first in the SM and then in
the MSSM.  Finally I discuss the analogous constraints and search
strategies for the superparticles, concentrating on the minimal
supergravity model.  Recent advances in identifying the most promising
channels for Higgs and SUSY searches are emphasised.

\newpage

As per the Standard Model (SM) the basic constituents of matter are
the quarks and leptons, which interact by the exchange of gauge bosons
-- photon, gluon and the massive $W$ and $Z$ bosons.  By now we have
seen all the quarks and leptons as well as the gauge bosons.  But the
story is not complete yet because of the mass and the hierarchy problems.
\bigskip

\noindent {\bf Mass Problem (Higgs Mechanism)}:
\medskip

The problem is how to give mass to the weak gauge bosons, $W$ and
$Z$, without breaking gauge symmetry, which is required for a
renormalisable field theory.  In order to appreciate it consider the
weak interaction Lagrangian of a charged scalar field $\phi$; i.e.
\be
{\cal L} = \left(\partial_\mu \phi + ig {\vec\tau \over 2} \vec W_\mu
\phi\right)^\dagger \left(\partial_\mu \phi + ig {\vec\tau \over 2}
\vec W_\mu \phi\right) - \left[\mu^2 \phi^\dagger \phi + \lambda
(\phi^\dagger \phi)^2\right] - {1 \over 4} \vec W_{\mu\nu} \vec
W_{\mu\nu}, 
\ee
where
\be
\vec W_{\mu\nu} = \partial_\mu \vec W_\nu - \partial_\nu \vec W_\mu -
g \vec W_\mu \times \vec W_\nu
\ee
is the field tensor for the weak gauge bosons $\vec W_\mu$.  The
charged and the neutral $W$ bosons form a $SU(2)$ vector, reflecting
the nonabelian nature of this gauge group.  This is responsbible for
the last term in (2), which leads to gauge boson
self-interaction.  Correspondingly the gauge transformation on $\vec
W_\mu$ has an extra term, i.e.
\be
\phi \rightarrow e^{i\vec \alpha \cdot \vec \tau} \phi, \ \vec W_\mu
\rightarrow \vec W_\mu - {1 \over g} \partial_\mu \vec\alpha -
\vec\alpha \times \vec W_\mu.
\ee
This ensures gauge invariance of $\vec W_{\mu\nu}$, and hence for the
last term of the Lagrangian, representing gauge kinetic energy.
Evidently the middle term, representing scalar mass and
self-interaction, is invariant under gauge transformation on $\phi$.
Finally the first term, representing scalar kinetic energy and gauge
interaction, can be easily shown to be invariant under the
simultaneous gauge transformations (3).  However the addition of a
mass term 
\be
- M^2 \vec W_\mu \cdot \vec W_\mu,
\ee
would clearly break the gauge invariance of the Lagrangian.  Note
that, in contrast the scalar mass term, $\mu^2 \phi^\dagger \phi$, is
clearly gauge invariant.  This phenomenon is exploited to give mass to
the gauge bosons through back door without breaking the gauge
invariance of the Lagrangian.  This is the celebrated Higgs mechanism
of spontaneous symmetry breaking [1].

One starts with a $SU(2)$ doublet of complex scalar field $\phi$ with
imaginary mass, i.e. $\mu^2 < 0$.  Consequently the minimum of the
scalar potential, $\mu^2 \phi^\dagger \phi + \lambda (\phi^\dagger
\phi)^2$, moves out from the origin to a finite value
\be
v = \sqrt{-\mu^2/\lambda},
\ee
i.e. the field develops a finite vacuum expectation value.  Since the 
perturbative expansion in quantum field theory is stable only around a
local minimum, one has to translate the field by the constant quantity,
\be
\phi^o = v + H^o(x),
\ee
where the superscript denotes the electric charge.  
Thus one gets a valid perturbative field theory in terms of the
redefined field $H$.  This represents the physical Higgs boson, while
the 3 other components of the complex doublet field are absorbed to
give mass and hence logitudinal components to the gauge bosons.

Substituting (6) in the first term of the Lagrangian (1) leads to a
mass term for $W$,
\be
M_W = {1\over2} g v.
\ee 
It also leads to a $HWW$ coupling,
\be
{1\over2} g^2 v = g M_W,
\ee
i.e. the Higgs coupling to the gauge bosons is propertional to the
gauge boson mass.  Similarly its couplings to quarks and leptons can
be shown to be propertional to their respective masses, i.e.
\be
h_{\ell,q} = m_{\ell,q}/v = {1\over2} g m_{\ell,q}/M_W.
\ee
Indeed, this is the source of the fermion masses in the SM.  Finally
substituting (6) in the middle term of the Lagrangian leads to a real
mass for the physical Higgs boson,
\be
M_H = v \sqrt{2\lambda} = M_W (2\sqrt{2\lambda}/g).
\ee
Substituting $M_W = 80 \GeV$ and $g = 0.65$ along with a perturbative
limit on the scalar self-coupling $\lambda \lsim 1$, implies that the
Higgs boson mass is bounded by $M_H < 1000 \GeV$.  But the story does
not end here.  Giving mass to the gauge bosons via the Higgs mechanism
leads to the so called hierarchy problem.
\bigskip

\noindent {\bf Hierarchy Problem (Supersymmetry)}:
\medskip

\nobreak
The problem is how to control the Higgs scalar mass in the desired 
range of a few hundered $\GeV$.  This is because the scalar masses are
known to have quadratically divergent quantum corrections from
radiative loops.  These would push
the output scalar mass to the cut-off scale of the SM, i.e. the GUT
scale $(10^{16} \GeV)$ or the Planck scale $(10^{19} \GeV)$.  The
desired mass range of $\sim 10^2 \GeV$ is clearly tiny compared to
these scales.  This is the so called hierarchy problem.  The
underlying reason for the quadratic divergence is that the scalar
masses are not protected by any 
symmetry unlike the fermion and the gauge boson masses, which are
protected by chiral symmetry and gauge symmetry.  Of course it was
this very property of the scalar mass that was exploited to give
masses to the fermions and gauge bosons in the first place.  The
hierarchy problem is the flip side of the same coin.

The most attractive solution to this problem is provided by
supersymmetry (SUSY), a symmetry between fermions and bosons [2].  It
predicts the quarks and leptons to have scalar superpartners called
squarks and sleptons $(\tilde q, \tilde \ell)$, and the gauge bosons
to have fermionic superpartners called gauginos $(\tilde g,\tilde \gamma,
\tilde W, \tilde Z)$.  In the minimal supersymmetric extension of the
standard model (MSSM) one needs two Higgs doublets $H_{1,2}$, with
opposite hypercharge $Y = \pm 1$, to give masses to the up and down
type quarks.  The corresponding fermionic superpartners are called
Higgsinos $(\tilde H_{1,2})$.  The opposite hypercharge of these two
sets of fermions ensures anomaly cancellation. 

SUSY ensures that the quadratically divergent quantum corrections from
quark, lepton and Higgs boson loops are cancelled by the contributions
from the corresponding squark, slepton and Higgsino loops.  Thus the
Higgs masses can be kept in the desired range of $\sim 10^2 \GeV$.
However this implies two important constraints on SUSY breaking.

\begin{enumerate}
\item[{i)}] SUSY can be broken in masses but not in couplings (soft
breaking), so that the co-efficients of the cancelling contributions
remain equal and opposite.
\item[{ii)}] The size of SUSY breaking in masses is $\sim 10^2 \GeV$,
so that the size of the remainder remains within this range.  Thus the
superpartners of the SM particles are also expected to lie in the mass
range of $\sim 10^2 \GeV$, going upto $1000 \GeV$.
\end{enumerate}
\bigskip

\noindent {\bf SM Higgs Boson}: {\bf Theoretical Constraints
\& Search Strategy}
\medskip

\nobreak
The Higgs self coupling $\lambda$ is ultra-violet divergent.  It
evolves according to the renormalisation group equation (RGE)
\be
{d\lambda \over d \ell n(Q/M_W)} = {3\lambda^2 \over 2\pi^2}.
\ee
It can be easily solved to give
\be
\lambda (Q) = {1 \over 1/\lambda (M_W) - (3/2\pi^2) \ell
n(Q/M_W)},
\ee
which has a Landau pole at 
\be
Q_0 = M_W e^{2\pi^2/3\lambda (M_W)}, \ \lambda (M_W) = {g^2 \over 8}
{M^2_H \over M^2_W}. 
\ee
Thus the larger the starting value $\lambda (M_W)$, the sooner will the
coupling diverge.  Evidently the
theory is valid only upto a cut-off scale $\wedge = Q_0$.
Requiring the theory to be valid at all energies, $\wedge \rightarrow
\infty$, would imply $\lambda (M_W) \rightarrow 0$; i.e. the only good
$\lambda \phi^4$ theory is a trivial theory.  Surely we do not want
that.  But if we want the theory to be valid upto the Planck scale or
GUT scale, we must have a relatively small $\lambda (M_W)$, which
corresponds to a small $M_H \lsim 200 \GeV$.  If on the other hand we
assume it to be valid only upto the TeV scale, then we can have a
larger $\lambda (M_W)$, corresponding to a relatively large $M_H \lsim
600 \GeV$.  This is the so-called triviality bound [3].  If $M_H$ is
significantly larger than $600 \GeV$, then the range of validity of
the theory is limited to $\wedge < 2 M_H$.  This would correspond to a
composite Higgs scenario, e.g. technicolour models.

Fig. 1 shows the triviality bound on the Higgs mass against the
cut-off scale $\wedge$ of the theory [4].  It also shows a lower bound
on the Higgs mass, which comes from a negative contribution to the RGE
(11) from the top Yukawa coupling, i.e.
\be
{d\lambda \over d \ell n (Q/M_W)} = {3 \over 2\pi^2} (\lambda^2 +
\lambda h^2_t - h^4_t).
\ee
The Yukawa coupling being ultra-violet divergent turns $\lambda$
negative at a high energy scale; and the smaller the starting value of
$\lambda$ (or equivalently $M_H$) the sooner will it become negative.
A negative $\lambda$ coupling has the undesirable feature of an
unstable vacuum.  Thus one can define a cut-off scale $\wedge$
for the theory, where this change of sign occurs.  The lower curve of
Fig. 1 shows the lower bound on $M_H$ as a function of the cut-off
scale $\wedge$ including the theoretical uncertainty [5].  We see from
this figure that the longer the range of validity of the theory, the
stronger will be the upper and lower bounds on $M_H$.  Thus assuming
no new physics upto the GUT or Planck scale (the desert scenario)
would constrain the SM Higgs mass to lie in the range
\be
M_H = 130 - 180 \GeV.
\ee
However the lower bound becomes invalid once we have more than one
Higgs doublet, since the unique relation between the top mass and
Yukawa coupling (9) will no longer hold.  In particular, one expects
an upper bound of $\sim 130 \GeV$ for the lightest Higgs boson mass in
MSSM instead of a lower bound, as we shall see below.  Since one
needs SUSY or some other form of new physics to stabilize the Higgs
mass, the above vacuum stability bound may have limited significance.
Nonetheless it is interesting to note that the predicted range of the
SM Higgs boson mass (15) agrees favourably with the indirect estimate
of this quantity from the precision measurement of electro-weak
parameters at LEP/SLD [6], i.e.
\be
M_H = 88^{+60}_{-37} \GeV \ \left(< 206 \GeV \ {\rm at} \ 95\% \
{\rm CL}\right).
\ee

The search strategy for Higgs boson is based on its preferential
coupling to the heavy quarks and gauge bosons as seen from (8,9).  The
LEP-I search was based on the so called Bjorken process
\be
e^+e^- \rightarrow Z \rightarrow HZ^\star \rightarrow \bar bb (\ell^+
\ell^-, \nu\nu, \bar qq),
\ee
while the LEP-II search is based on the associated process with $Z$
and $Z^\star$ intercharged, resulting in the limit [7] 
\be
M_H > 114.1 \GeV.
\ee

Thus a promising mass range to probe for the SM Higgs boson signals is
\be
M_H = 114 - 206 \GeV.
\ee

But the upper limit is not a robust one since the underlying quantum
corrections have only logarithmic dependence on the Higgs mass. 
Fig. 2 shows the total decay width of the Higgs boson 
along with the branching ratios for the important decay channels [8].
It is clear from this figure that the mass range can be devided into
two parts -- a) $M_H < 2 M_W (90 - 160 \GeV)$ and b) $M_H > 2M_W (160
- 1000 \GeV)$.

The first part is the so called intermediate mass region, where the
Higgs width is expected to be only a few MeV.  The dominant decay mode
is $H \rightarrow \bar bb$.  This has unfortunately a huge QCD
background, which is $\sim 1000$ times larger than the signal.  By far
the cleanest channel is $\gamma\gamma$, where the continuum background
is a 2nd order $EW$ process.  However, it suffers from a small
branching ratio 
\be
B (H \rightarrow \gamma\gamma) \sim 1/1000,
\ee
since it is a higher order process, induced by the $W$ boson loop.  So
one needs a very high jet/$\gamma$ rejection factor $\gsim 10^4$.
Besides the continuum background being propertional to $\Delta
M_{\gamma\gamma}$, one needs a high resolution,
\be
\Delta M_{\gamma\gamma} \lsim 1 \GeV \ {\rm i.e.} \ \lsim 1\% \ {\rm
of} \ M_H.
\ee
This requires fine $EM$ calorimetry, capable of measuring the $\gamma$
energy and direction to $1\%$ accuracy.  

One can get a feel for the size of the signal from the Higgs
production cross-sections shown in Fig. 3.  The relevant production
processes are 
\be 
gg \ {\buildrel {\bar t^\star t^\star} \over
\longrightarrow} \ H, 
\ee 
\be qq \ {\buildrel {W^\star W^\star} \over
\longrightarrow} \ H q q, 
\ee 
\be q\bar q' \ {\buildrel {W^\star}
\over \rightarrow} \ HW, 
\ee 
\be gg, q\bar q \rightarrow Ht \bar t (Hb \bar b).  
\ee 
The largest cross-section, coming from gluon-gluon
fusion via the top quark loop (22), is of the order of $10 pb$ in the
intermediate mass region.  Thus the expected size of the $H
\rightarrow \gamma\gamma$ signal is $\sim 10 fb$, corresponding to
$\sim 10^3$ events at the high luminosity ($\sim$ 100 fb$^{-1}$) run
of LHC.  The estimated continuum background is $\sim 10^4$ events,
which can of course be subtracted out.  Thus the significance of the
signal is given by its relative size with respect to the statistical
uncertainty in the background, i.e.  
\be 
S/\sqrt{B} \simeq 10.  
\ee 
Fig. 4 shows the ATLAS (left) and CMS (right) simulations for the SM
Higgs signals at LHC from different decay channels.  The ATLAS figure
shows the expected significance level of the signal for the high
luminosity (100 fb$^{-1}$) run of LHC.  Combining the different decay
channels should give a $\geq 10 \sigma$ signal over the entire Higgs
mass range of 100 - 1000 GeV.  The CMS figure shows the minimum
luminosity required for the discovery of a $5\sigma$ Higgs signal
against its mass.  It shows that a modest luminosity of $\sim 20$
fb$^{-1}$, which is expected to be accumulated at the low luminosity
run, should suffice for this discovery over most of Higgs mass range
of interest.

As we see from Fig. 4, the most promising Higgs decay channel is
\be
H \rightarrow ZZ \rightarrow \ell^+ \ell^- \ell^+ \ell^-,
\ee
since reconstruction of the $\ell^+ \ell^-$ invariant masses makes it
practically background free.  Thus it provides the most important
Higgs signal right from the subthresold region of $M_H = 140 \GeV$
upto $600 \GeV$.  Note however a sharp dip in the $ZZ$
branching ratio at $M_H = 160 - 170 \GeV$ due to the opening of the
$WW$ channel (see Fig. 2).  The most important Higgs signal in this
dip region is expected to come from [10]
\be
H \rightarrow WW \rightarrow \ell^+ \nu \ell^- \bar\nu.
\ee
However in general this channel suffers from a much larger background
for two reasons -- i) it is not possible to reconstruct the $W$ masses
because of the two neutrinos and ii) there is a large $WW$ background
from $t\bar t$ decay.

For large Higgs mass, $M_H = 600 - 1000 \GeV$, the 4-lepton signal
(27) becomes too small in size.  In this case the decay channels 
\be
H \rightarrow WW \rightarrow \ell \nu q \bar q', \ \ H \rightarrow ZZ
\rightarrow \ell^+ \ell^- \nu \nu
\ee
are expected to provide more favourable signals.  The biggest
background comes from single $W(Z)$ production along with QCD jets.
However, one can exploit the fact that a large part of the signal
cross-section in this case comes from $WW$ fusion (23), which is
accompanied by two forward (large-rapidity) jets.  One can use the
double forward jet tagging to effectively control the background.
Indeed the above simulation studies by the CMS and ATLAS collaborations show
that using this strategy one can extend the Higgs search right upto
$1000 \GeV$ [9].
\bigskip

\noindent {\bf MSSM Higgs Bosons: Theoretical Constraints \&
Search Strategy}
\medskip

\nobreak
As mentioned earlier, the MSSM contains two Higgs doublets, which
correspond to 8 independent states.  After 3 of them are absorbed by the
$W$ and $Z$ bosons, one is left with 5 physical states: two neutral
scalars $h^0$ and $H^0$, a pseudoscalar $A^0$, and a pair of charged
Higgs scalars $H^\pm$.  At the tree-level their masses and couplings
are determined by only two parameters -- the ratio of the two vacuum
expectation values, $\tan\beta$, and one of the scalar masses, usually
taken to be $M_A$.  However, the neutral scalars get a large radiative
correction from the top quark loop along with the top squark (stop)
loop.  To a good approximation this is given by [11]
\be
\epsilon = {3g^2 m^4_t \over 8\pi^2 M^2_W} \ell n\left({M^2_{\tilde
t} \over m^2_t}\right),
\ee
plus an additional contribution from the $\tilde t_{L,R}$ mixing,
\be
\epsilon_{\rm mix} = {3g^2 m^4_t \over 8\pi^2 M^2_W} {A^2_t \over
M^2_{\tilde t}} \left(1 - {A^2_t \over 12 M^2_{\tilde t}}\right) \leq
{9g^2 m^4_t \over 8\pi^2 M^2_W}.
\ee
Thus while the size of $\epsilon_{\rm mix}$ depends on the trilinear
SUSY breaking parameter $A_t$, it has a definite maximum value.  As
expected the radiative corrections vanish in the exact SUSY 
limit.  One can estimate the rough magnitude of these corrections
assuming a SUSY breaking scale of $M_{\tilde t} = 1$ TeV.  The leading
log QCD corrections can be taken into account by using the running
mass of top at the appropriate energy scale [11]; i.e. $m_t (\sqrt{m_t
M_{\tilde t}}) \simeq 157 \GeV$ in (30) and $m_t (M_{\tilde t}) \simeq
150 \GeV$ in (31) instead of the top pole mass of $175 \GeV$.  One
can easily check the resulting size of the radiative corrections are 
\be
\epsilon \sim M^2_W \ \ {\rm and} \ \ 0 < \epsilon_{\rm mix} \lsim
M^2_W. 
\ee

The neutral scalar masses are obtained by diagonalising the
mass-squared matrix
\be
\left(\matrix{M^2_A \sin^2 \beta + M^2_Z \cos^2 \beta & -(M^2_A +
M^2_Z) \sin \beta \cos \beta \cr & \cr -(M^2_A + M^2_Z) \sin \beta
\cos \beta & M^2_A \cos^2 \beta + M^2_Z \sin^2 \beta +
\epsilon'}\right)
\ee
with $\epsilon' = (\epsilon + \epsilon_{\rm mix})/\sin^2 \beta$.  Thus
\beq
M^2_h &=& {1\over2} \Bigg[M^2_A + M^2_Z + \epsilon' - \Big\{(M^2_A +
M^2_Z + \epsilon')^2 - 4M^2_A M^2_Z \cos^2 \beta \nonumber \\[2mm]
& & ~~~~ - 4\epsilon' (M^2_A \sin^2 \beta + M^2_Z \cos^2
\beta)\Big\}^{1/2}\Bigg] \nonumber \\[2mm]
M^2_H &=& M^2_A + M^2_Z + \epsilon' - M^2_h \nonumber \\[2mm]
M^2_{H^\pm} &=& M^2_A + M^2_W
\eeq
where $h$ denotes the lighter neutral scalar [12].  One can easily
check that its mass has an asymptotic limit for $M_A \gg M_Z$, i.e.
\be
M^2_h \longrightarrow M^2_Z \cos^2 2\beta + \epsilon + \epsilon_{\rm
mix}, 
\ee
while $M^2_H$, $M^2_{H^\pm} \rightarrow M^2_A$.  Thus the MSSM
contains at least one light Higgs boson $h$, whose tree-level mass
limit $M_h < M_Z$, goes upto $130 \GeV$ after including the
radiative corrections.  

Let us consider now the couplings of the MSSM Higgs bosons.  A
convenient parameter for this purpose is the mixing angle $\alpha$
between the neutral scalars, i.e.
\be
\tan 2\alpha = \tan 2\beta {M^2_A + M^2_Z \over M^2_A - M^2_Z +
\epsilon'/\cos 2\beta}, -\pi/2 < \alpha < 0.
\ee
Note that
\be
\alpha \ {\buildrel {M_A \gg M_Z} \over \longrightarrow} \ \beta -
\pi/2. 
\ee
\begin{enumerate}
\item[{}] Table-I.  Important couplings of the MSSM Higgs bosons $h$,
$H$ and $A$ relative to those of the SM Higgs boson
\end{enumerate}
\[
\begin{tabular}{|c|c|c|c|c|}
\hline
&&&& \\
Channel & $H_{\rm SM}$ & $h$ & $H$ & $A$ \\
&&&& \\
\hline
&&&& \\
$\bar bb(\tau^+\tau^-)$ & $\displaystyle{gm_b \over 2M_W} (m_\tau)$ & $-\sin
\alpha/\cos \beta$ & $\cos \alpha/\cos \beta$ & $\tan \beta$ \\
 & & $\rightarrow 1$ & $\tan \beta$ & '' \\
&&&& \\
\hline
&&&& \\
$\bar tt$ & $\displaystyle g{m_t \over 2M_W}$ & $\cos\alpha/\sin\beta$ &
$\sin\alpha/\sin\beta$ & $\cot \beta$ \\
& & $\rightarrow 1$ & $\cot\beta$ & '' \\
&&&& \\
\hline
&&&& \\
$WW (ZZ)$ & $g M_W (M_Z)$ & $\sin (\beta - \alpha)$ & $\cos (\beta -
\alpha)$ & $0$ \\
& & $\rightarrow 1$ & $0$ & '' \\
&&&& \\
\hline
\end{tabular}
\]

\noindent Table-I shows the important couplings of the neutral Higgs bosons
relative to those of the SM Higgs boson.  The limiting values of these
couplings at large $M_A$ are indicated by arrows.  The corresponding
couplings of the charged Higgs boson, which has no SM analogue, are 
\beq
H^+ \bar t b &:& {g \over \sqrt{2}M_W} (m_t \cot \beta + m_b \tan
\beta), \ H^+ \tau \nu : {g \over \sqrt{2}M_W} m_\tau \tan \beta,
\nonumber \\[2mm]
H^+ W^- Z &:& 0.
\eeq
Note that the top Yukawa coupling is ultraviolet divergent.  Assuming
it to lie within the perturbation theory limit all the way upto the
GUT scale implies 
\be
1 < \tan \beta < m_t/m_b,
\ee
which is therefore the favoured range of $\tan\beta$.  However, it
assumes no new physics beyond the MSSM upto the GUT scale, which is a
stronger assumption than MSSM itself.  Nontheless we shall concentrate
in this range.

Before discussing the search of MSSM Higgs bosons at LHC let us
briefly discuss the LEP constraints on these particles.  Fig. 5 plots
the $h^0$, $H^0$ and $H^\pm$ masses against $M_A$ for two
representative values of $\tan\beta$ ($= 3$ and 30) assuming maximum
stop mixing [13].  It also plots the corresponding $\sin^2$ $(\beta -
\alpha)$, representing the suppression factor of the $h$ signal
relative to the SM Higgs boson for the LEP process (17).  We see that
for $\tan\beta = 3$, the maximum value of $h$ mass is marginally above
the SM Higgs mass limit of 114 GeV.  Moreover the corresponding lower
limit of $M_h$ is marginally smaller than this value since the signal
suppression factor is $\geq 0.5$.  Thus $\tan\beta = 3$ lies just
inside the LEP allowed region, while it disallows $\tan\beta \leq
2.4$.  The disallowed region extends over $\tan\beta \lsim 5$ for a
more typical value of the mixing parameter, $A_t \simeq 1$ TeV.  One
also sees from this figure that the lower limit of $M_h \geq 114$ GeV
will hold at 
large $\tan\beta$ ($\sim 30)$ if $M_A$ is $> 130$ GeV.  But for lower
values of $M_A$ this signal is strongly suppressed at large
$\tan\beta$; and one can only get a modest limit of $M_A \simeq M_h >
90$ GeV from the pair-production process $e^+e^- \rightarrow hA$ at
LEP [7].  The pair production of charged Higgs bosons at LEP gives a
limit $M_{H^\pm} > 78$ GeV, which is close to its theoretical mass
limit (34).

Coming back to the neutral Higgs couplings of Table-I, we see that in
the large $M_A$ limit the light Higgs boson $(h)$ couplings approach
the SM values.  The other Higgs bosons are not only heavy, but their
most important couplings are also suppressed.  This is the so
called decoupling limit, where the MSSM Higgs sector is
phenomenologically indistinguishable from the SM.  It follows
therefore that the Higgs search stategy at LHC for $M_A \gg M_Z$ should be
the same as the SM case, i.e. via 
\be
h \rightarrow \gamma\gamma.
\ee

At lower $M_A$, several of the MSSM Higgs bosons become light.
Unfortunately their couplings to the most important channels, $\bar
tt$ and $WW/ZZ$, are suppressed relative to the SM Higgs boson [12].
Thus their most important production cross-sections as well as their
decay BRs into the $\gamma\gamma$ channel are suppressed relative to
the SM case.  Consequently the Higgs detection in this region is very
challenging.  Nonetheless recent simulation studies show that it will
be possible to see at least one of the MSSM Higgs bosons at LHC over
the full parameter space of $M_A$ and $\tan\beta$.  Fig. 6 shows such
a simulation by the CMS collaboration [14] for integrated luminosities
of 30 fb$^{-1}$ and 100 fb$^{-1}$, which are expected from the low and
high luminosity runs of LHC respectively.  It looks much more
promising now than 4-5 years back, when the corresponding plot showed
a big hole in the middle of this parameter space [15].  The
improvement comes from the following three processes, which have been
studied only during the last few years.

\begin{enumerate}
\item[{1)}] $t\bar t h$, $h \rightarrow b\bar b$: The $h \rightarrow
b\bar b$ decay width is enhanced by the $\sin^2 \alpha/\cos^2 \beta$
factor, while the $h \rightarrow \gamma\gamma$ width via the $W$ boson
loop is suppressed by $\sin^2 (\beta - \alpha)$ as $M_A \rightarrow
M_Z$.  Besides in the latter case the production cross-section via the
top quark loop is suppressed by a cancelling contribution from the
stop loop.  Hence the above process provides a viable signature for
$h$ over the modest $m_A$ region where the canonical $h \rightarrow
\gamma\gamma$ signature becomes too small.

\item[{2)}] $tH^\pm$, $H^\pm \rightarrow \tau\nu (tb)$: While the
earlier analyses of charged Higgs boson signal at LHC were restricted
to $M_{H^\pm} < m_t$ ($M_A \lsim 140$ GeV) [16], recently they have
been extended for heavier $H^\pm$ via these processes [17,18].  In
particular the associated production of $t H^\pm$ followed by the
$H^\pm \rightarrow \tau\nu$ decay is seen to provide a viable
signature over a large range of $M_{H^\pm}$ $(M_A)$ for $\tan\beta
\gsim 10$.  Here one exploits the predicted $\tau$ polarization, i.e. 
$P_\tau = +1$ for the $H^\pm$ signal and -1 for the $W^\pm$ background.  In
the 1-prong hadronic decay channel of $\tau$, the $P_\tau = +1 (-1)$ state is
peaked at $R \simeq 1 \ (0.4)$, where $R$ denotes the fraction of the
visible $\tau$-jet momentum carried by the charged prong [18].
Following this suggestion a simple kinematic cut of $R > 0.8$ has been used in
the above simulation [14] to effectively suppresses the $W^\pm
\rightarrow \tau\nu$ as well as the fake $\tau$ background from QCD
jets, while retaining nearly half the signal events.

\item[{3)}] $H,A \rightarrow \tau^+ \tau^- \rightarrow 2\tau$-jets:
Earlier analyses of this process assumed at least one of the $\tau$'s
to have leptonic decay.  The above simulation shows that hadronic
decay of both the taus provides a viable signature over a wider range
of $M_A$.  This signature can be improved further by exploiting the
correlation between the polarizations of the 2 taus as suggested in
[19].
\end{enumerate}

Note that all these three new channels require identification of $b$
quark and/or hadronic $\tau$-jet.  Thus they are based on tracker
performance, while the canonical $h \rightarrow \gamma\gamma$ channel
emphasised EM calorimeter.

Finally one should note from Fig. 6 that there is a large part of the
parameter space, where one can see only one Higgs boson $(h)$ with SM
like couplings and hence not be able to distinguish the SUSY Higgs
sector from the SM.  Fortunately it will be possible to probe SUSY
directly via superparticle search at LHC as we see below.  
\bigskip

\noindent {\bf Superparticles: Signature \& Search Strategy}
\medskip

\nobreak
I shall concentrate on the standard $R$-parity conserving SUSY model,
where 
\be
R = (-1)^{3B+L+2S}
\ee
is defined to be $+1$ for the SM particles and $-1$ for their
superpartners, since they differ by $1/2$ unit of spin $S$.  It
automatically ensures Lepton and Baryon number conservation by
preventing single emission (absorption) of superparticle.

Thus $R$-conservation implies that (i) superparticles are produced in
pair and (ii) the lightest superparticle (LSP) is stable.
There are strong astrophysical consrtaints against such a stable
particle carrying colour 
or electric charge, which imply that the LSP is either sneutrino $\tilde \nu$
or photino $\tilde\gamma$ (or in general the lightest neutralino).
The latter alternative is favoured by most SUSY models.  In
either case the LSP is expected to have only weak interaction with
ordinary matter like the neutrino, since e.g.
\be
\tilde\gamma q \ {\buildrel {\tilde q} \over \longrightarrow} \ q
\tilde \gamma \ \ {\rm and} \ \ \nu q \ {\buildrel W \over
\longrightarrow} \ e q'
\ee
have both electroweak couplings and $M_{\tilde q} \sim M_W$.  This
makes the LSP an ideal candidate for the Cold Dark Matter.  It also
implies that the LSP would leave the normal detectors without a trace
like the neutrino.  The resulting imbalance in the visible momentum
constitutes the canonical missing transverse-momentum $(p\!\!\!/_T)$
signature for superparticle production at hadron colliders.  It is
also called the missing transverse-energy $(E\!\!\!/_T)$ as it is
often measured as a vector sum of the calorimetric energy deposits in
the transverse plane.

The main processes of superparticle production at LHC are the QCD
processes of quark-antiquark and gluon-gluon fusion [20]
\be
q\bar q, gg \longrightarrow \tilde q \bar{\tilde q} (\tilde g \tilde
g).
\ee
The NLO corrections can increase these cross-sections by $15-20\%$
[21].  The simplest decay processes for the produced squarks and
gluinos are
\be
\tilde q \rightarrow q \tilde \gamma, \ \tilde g \rightarrow q\bar q
\tilde \gamma.
\ee
Convoluting these with the pair production cross-sections gave 
the simplest jets + $p\!\!\!/_T$ signature for squark/gluino
production, which were adequate for the early searches for relatively
light squarks and gluinos.  However, over the mass range of current
interest $(\geq 100 \GeV)$ the cascade decays of squark and gluino
into the LSP via the heavier chargino/neutralino states are expected
to dominate over the direct decays.  This is both good news and
bad news.  On the one hand the cascade decay degrades the missing-$p_T$
of the canonical jets $+ p\!\!\!/_T$ signature [22].  But on the other hand
it gives a new multilepton signature via the leptonic decays of these
chargino/neutralino states [23].  It may be noted here that one gets a mass
limit of $M_{\tilde q,\tilde g} \gsim 200 \GeV$ from the Tevatron data
using either of the two signatures [7]. 

The cascade decay is described in terms of the $SU(2) \times U(1)$
gauginos $\tilde W^{\pm,0}, \tilde B^0$ along with the Higgsinos
$\tilde H^\pm$, $\tilde H^0_1$ and $\tilde H^0_2$.  The $\tilde B$ and
$\tilde W$ masses are denoted by $M_1$ and $M_2$ respectively while
the Higgsino masses are given by the supersymmetric mass
parameter $\mu$.  The charged and the neutral gauginos
will mix with the corresponding Higgsinos to give the physical
chargino $\chi^\pm_{1,2}$ and neutralino $\chi^0_{1,2,3,4}$ states.
Their masses and compositions can be found by diagonalising the
corresponding mass matrices, i.e.
\be
M_C = \left(\matrix{M_2 & \sqrt{2} M_W \sin \beta \cr & \cr \sqrt{2}
M_W \cos \beta & \mu}\right),
\ee
\bigskip
{\small
\[
M_N = \left(\matrix{M_1 & 0 & -M_Z \sin \theta_W \cos \beta & M_Z \sin
\theta_W \sin \beta \cr & & & \cr 0 & M_2 & M_Z \cos \theta_W \cos
\beta & -M_Z \cos \theta_W \sin \beta \cr & & & \cr -M_Z \sin \theta_W
\cos \beta & M_Z \cos \theta_W \cos \beta & 0 & -\mu \cr & & & \cr M_Z
\sin \theta_W \sin \beta & -M_Z \cos \theta_W \sin \beta & -\mu &
0}\right).
\]
}
\be
\ee

The LEP limit [7] on the lighter chargino $\chi^\pm_1$ mass is 100
GeV, which implies
\be
|\mu|,|M_2| > 100 \GeV.
\ee
The corresponding slepton mass limits are $m_{\tilde e} > 99 \GeV$,
$m_{\tilde\mu} > 95 \GeV$ and $m_{\tilde\tau} > 80 \GeV$.  The
sneutrino $\tilde\nu$ and the lightest neutralino $\chi^0_1$ mass
limits are 45 and 40 GeV respectively.  In general the cascade decay
of squarks and gluinos would depend on all these masses.
\bigskip

\noindent {\bf SUGRA Model:} To control the number of mass parameters one
has to assume a supersymmetry breaking model.  The simplest and most
popular model is called supergravity, where SUSY is broken in a hidden
sector and its effect is communicated to the observable sector via
gravitational interaction.  Since this interaction is colour and
flavour blind, it leads to a common SUSY breaking mass for all the
scalars $(m_0)$ and another one for all the gauginos $(M_{1/2})$ near
the GUT scale.  This is consistent with the successful unification of
the $SU(3) \times SU(2) \times U(1)$ gauge couplings at this scale
[7].  Then the SUSY breaking masses evolve to low energy scales as per
the Renormalisation Group Evolution formulae [24].

The gaugino masses evolve like the corresponding gauge couplings, i.e.
\be
M_i (Q) = M_{1/2} \alpha_i (Q)/\alpha (M_G).
\ee
Thus at the low energy scale, $Q \sim M_W$,
\beq
M_2 &=& M_{1/2} \alpha_2/\alpha(M_G) \simeq 0.8 M_{1/2},
\nonumber\\[2mm] 
M_1 &=& M_2 \alpha_1/\alpha_2 \simeq M_2/2, \nonumber \\[2mm] 
M_{\tilde g} &=& M_3 = M_2 \alpha_3/\alpha_2 \simeq 3 M_2.
\eeq

Ignoring the trilinear coupling $(A)$ terms, one can write the SUSY
breaking scalar masses at low energy as 
\be
m^2_i = m^2_0 + a_i m^2_0 + b_i M^2_{1/2},
\ee
where $a_i$ is proportional to Yukawa coupling and $b_i$ to
combination of gauge and Yukawa couplings.  Of course the Yukawa
couplings are significant only for $H_2$ and the 3rd generation
squarks and sleptons.  It drives the $H_2$ mass square $(m^2_2 =
m^2_{H_2} + \mu^2)$ negative, as required for EWSB.  The EWSB
condition is
\be
{M^2_Z \over 2} = {m^2_{H_1} - m^2_{H_2} \tan^2\beta \over \tan^2\beta
- 1} - \mu^2,
\ee
which reduce to $-m^2_{H_2} - \mu^2$ over the range $\tan\beta \gsim
5$, favoured by LEP.  Substituting the evolution eq. (50) for
$m^2_{H_{1,2}}$ in (51) gives [24]
\be
\mu^2 \simeq m^2_0 \left({9 \over 7} y-1\right) - M^2_{1/2} \left(0.5
- 6y + {18 \over 7} y^2\right) - {M^2_Z \over 2},
\ee
where $y$ denotes the top Yukawa coupling relative to its fixed point
value.  For the physical top mass of 175 GeV it is given by [25,26,27]
\be
y = {h^2_t \over h^2_f} = {1 + 1/\tan^2\beta \over 1.44} \simeq 0.71,
\ee
where the last equality holds to within 2\% accuracy over $\tan\beta
\geq 5$.  Subtituting this in (52) give
\be
\mu^2 \simeq -0.08 m^2_0 + 2.4 M^2_{1/2} - 0.5 M^2_Z.
\ee
Thus one has only two independent parameters $m_0$ and $M_{1/2}$ apart
from $\tan\beta$ and the sign of $\mu$.  The small coefficient of the
$m^2_0$ term in (54) along with (49) imply that over most of the
parameter space $\mu^2 > M^2_2$, i.e. the lighter chargino and
neutralino states are gauginos $(\tilde W,\tilde B)$ obeying the mass
hierarchy (49).  However there is a narrow strip of very high $m^2_0$
region, where its negative contribution pushes $|\mu|$ down to the LEP
limit of 100 GeV.  Here the 
lighter chargino and neutralinos are Higgsino or mixed states.  This
is the focus point region of ref. [26], which is favoured by the
electron and neutron EDM constraints [28].  It is also favoured by the
cosmological constraint on the relic density of Dark Matter [29], as
we see below.

Fig. 7 shows the contours of DM relic density in the $m_0 - M_{1/2}$
plane for a representative value of $\tan\beta = 10$ and +ve $\mu$
[30].  The latter is favoured by the indirect constraint from $b
\rightarrow s\gamma$.  The regions marked I and II correspond to $\mu
< 100 \GeV$ and $m_{\tilde\tau_1} < m_{\tilde\chi^0_1}$ respectively.
The former is excluded by the LEP constraint (47) and the latter by
the requirement of a neutral LSP.  The remaining area is within the
discovery limit of LHC.  The indicated upper limit from the muon 
anomalous magnetic moment data is not compelling because of the
uncertainty in the QCD contribution [31].  More importantly however
even a bigger region is excluded by the cosmological constraint of DM
relic density [7]
\be
0.1 < \Omega h^2 < 0.3.
\ee
While the lower limit may be evaded by assuming alternative DM
candidates, the upper limit is quite compelling.

The reason for the predicted over abundance of SUSY DM over most of
the parameter space is that the LSP $(\chi^0_1)$ over this region is
dominantly $\tilde B$, which does not couple to $W$ or $Z$ bosons.
Thus they can pair-annihilate only via the exchange of massive
superparticles like squarks or sleptons, $\tilde B\tilde B {\buildrel
{\tilde q(\tilde\ell)} \over \longrightarrow} q\bar q (\ell^+
\ell^-)$, which have low rates.  However there are two strips
adjacent to the disallowed regions I and II, which predict right DM
relic densities.  In the first one the LSP has a large Higgsino
component, so that it can pair annihilate by Higgsino exchange $\tilde
H^0 \tilde H^0 {\buildrel {\tilde H^0 (\tilde H^\pm)} \over
\longrightarrow} ZZ (W^+ W^-)$ or simply by $s$-channel $Z$ exchange
$\tilde H^0 \tilde H^0 {\buildrel Z \over \rightarrow} q\bar q$.  At
the boundary of the excluded region I the lighter chargino and
neutralino states are dominantly Higgsinos, $\tilde H^\pm$ and $\tilde
H^0_{1,2}$, with nearly degenerate mass $\simeq \mu$.  Thus there is
also a large co-annihilation rate via $s$-channel $W$ exchange $\tilde
H^0 \tilde H^\pm {\buildrel W^\pm \over \rightarrow} q' \bar q$.  This
leads to underabundance of SUSY DM $(\Omega h^2 < 0.1)$ at this
boundary.  In the second region the $\tilde\tau_1$ mass is close to
that of the LSP $(\tilde B)$.  Thus one has a fairly large
pair-annihilation rate via $\tilde\tau_1$ exchange, resulting in the
desired relic density (52).  There is also co-annihilation of
$\tilde\tau_1$ with $\tilde B$ via $s$-channel $\tau$ at the
boundary.  Similar features hold at other values of $\tan\beta$ as
well as negative $\mu$ [31].

Thus there is a good deal of current interest in the LHC signature of
superparticles in these two strips, corresponding to i) $m_0 \gg
M_{1/2}$ and ii) $m_0 \ll M_{1/2}$.

\begin{enumerate}
\item[{i)}] This corresponds to the above mentioned focus point
region.  A distinctive feature of this region is an inverted mass
hierarchy, where the top squarks $(\tilde t_{1,2})$ are predicted to
be significantly lighter than those of 1st two generations.  This is
because $\tilde t_{L,R}$ have large negative Yukawa coupling
contributions $(a_i)$ in eq. (50), while the 1st two generation squark
masses are $\simeq m_0$.  Thus for $m_0 = 2000 \GeV$, $M_{1/2} = 500
\GeV$ and $\tan\beta = 10$ one predicts [27]
\be
M_{\tilde g} \simeq 1300 \GeV, \ m_{\tilde t_1} \simeq 1500 \GeV, \
m_{\tilde u,\tilde d} \geq 2200 \GeV.
\ee
Consequently one predicts a large branching fraction for gluino decay
via $\tilde t_1$, i.e.
\be
\tilde g {\buildrel \tilde t_1 \over \rightarrow} \bar t t
\tilde\chi^0_i, \ \bar t b \chi^\pm_j \rightarrow 2b 2W \chi^0_1
\cdots .
\ee
The corresponding final state from gluino pair production contains
$4b$ and $4W$ particles.  Fig. 8 shows the resulting signal in single
lepton, dilepton, same-sign dilepton and trilepton channels along with
$4b$ tags [27].  The background is effectively suppressed by a 100 GeV
cut on the acconpanying ${E\!\!\!/}_T$.  The large multiplicity of
$b$-quarks and $W$ bosons makes this a far more spectacular signal
compared to the standard cascade decade case.

\item[{ii)}] In this region the $\tilde\tau_1$ mass is close that of
$\tilde\chi^0_1$.  Consequently there is a large branching fraction of
cascade decay via $\tilde\tau_1$ into 
\be
\tilde\tau_1 \rightarrow \tau \chi^0_1.
\ee
Thus the final state contains two $\tau$'s along with a large
${E\!\!\!/}_T$.  In this case the most promising signature corresponds
to 1-prong hadronic decay of the $\tau$'s [32].  A remarkable
prediction of the SUGRA model is the polarization of $\tau$ coming
from the $\tilde\tau_1$ decay (58).  Fig. 9 shows that $P_\tau > 0.9$
over the full $m_0 - M_{1/2}$ plane at $\tan\beta = 30$, which holds
at other values of $\tan\beta$ as well [33].  Moreover $P_\tau > 0.95$
in the relevant half-plane of $m_0 < M_{1/2}$.  In contrast the SM
background from $W$ and $Z$ decays correspond to $P_\tau = -1$ and 0
respectively.  Fig. 10 shows the $R$ distributions for $P_\tau =
+1,0,-1$.  As in the case of $H^\pm$ signal discussed earlier, one can
also sharpen the SUSY signal by demanding $R > 0.8$ -- i.e. the
charged prong to carry $> 80\%$ of the visible $\tau$-jet momentum.
\end{enumerate}

Note that the SUSY signals in the above two regions are based on
identification of $b$ and hadronic $\tau$-jets.  Thus they again
empasize the tracker performance like the MSSM Higgs signals discussed
earlier. 

Jan Kwiecinski is not only an old collaborator but a very good family
friend of ours since the early seventies. It gives me great pleasure
to dedicate this article to him on his 65th birth anniversary.

\newpage

\noindent {\bf References}:
\medskip

\begin{enumerate}
\item[{1.}] For a review see e.g. J.F. Gunion, H.E. Haber, G. Kane and
S. Dawson, The Higgs Hunters' Guide (Addison-Wesley, Reading, MA,
1990). 
\item[{2.}] For a review see e.g. H.E. Haber and G. Kane,
Phys. Rep. 117, 75 (1985). 
\item[{3.}] L. Maiani, G. Parisi and R. Petronzio, Nucl. Phys. B136,
115 (1978); N. Cabbibo, L. Maiani, G. Parisi and R. Petronzio,
Nucl. Phys. B158, 295 (1979); M. Lindner, Z. Phys. C31, 295 (1986).
\item[{4.}] T. Hambye and K. Riesselmann, Phys. Rev. D55, 7255 (1997). 
\item[{5.}] G. Altarelli and G. Isidori, Phys. Lett. B337, 141 (1994);
J.A. Casas, J.R. Espinosa and M. Quiros, Phys. Lett. B342, 171 (1995)
and B382, 374 (1996). 
\item[{6.}] LEP and SLC Electroweak Working Group, hep-ex/0112021.
\item[{7.}] Review of Particle Properties: K. Hagiwara et al,
Phys. REv. D66, 010001 (2002).
\item[{8.}] M. Spira, Fortsch. Phys. 46, 203 (1998).
\item[{9.}] ATLAS Collaboration: Technical Design Report,
CERN/LHCC/99-15 (1999); CMS Collaboration: M. Dittmar, Pramana 55, 151
(2000); D. Dittmar and A.S. Nicollerat, CMS-NOTE 2001/036.
\item[{10.}] M. Dittmar and H. Dreiner, Phys. Rev. D55, 167 (1997). 
\item[{11.}] For a  review of radiative correction along with
reference to the earlier works see H.E. Haber, R. Hempfling and
A.H. Hoang, Z. Phys. C75, 539 (1997).
\item[{12.}] See e.g. A. Djouadi, J. Kalinowski and P.M. Zerwas,
Z. Phys. C70, 435 (1996).
\item[{13.}] A. Djouadi [hep-ph/0205248].
\item[{14.}] D. Denegri et al, CMS NOTE 2001/032 [hep-ph/0112045].
\item[{15.}] D.P. Roy, Pramana 51, 7 (1998) [hep-ph/9803421], see
Fig. 7.
\item[{16.}] D.P. Roy, Phys. Lett. B277, 183 (1992) and
Phys. Lett. B283, 403 (1992); S. Raychaudhuri and D.P. Roy,
Phys. Rev. D53, 4902 (1996).  
\item[{17.}] S. Moretti and D.P. Roy, Phys. Lett. B470, 209 (1999);
D. Miller, S. Moretti, D.P. Roy and W. Sterling, Phys. Rev. D61,
055011 (2000).
\item[{18.}] D.P. Roy, Phys. Lett. B459, 607 (1999).
\item[{19.}] S. Moretti and D.P. Roy, Phys. Lett. B545, 329 (2002).
\item[{20.}] G.L. Kane and J.P. Leville, Phys. Lett. B112, 227 (1982);
P.R. Harrison and C.H. Llewellyn-Smith, Nucl. Phys. B213, 223 (1983)
[Err. Nucl. Phys. B223, 542 (1983)]; E. Reya and D.P. Roy,
Phys. Lett. B141, 442 (1984); Phys. Rev. D32, 645 (1985).
\item[{21.}] W. Beenakker, R. Hopker, M. Spira and P. Zerwas,
Nucl. Phys. B492, 51 (1997); M. Kramer, T. Plehn, M. Spira and
P. Zerwas, Phys. Rev. Lett. 79, 341 (1997).
\item[{22.}] H. Baer, C. Chen, F. Paige and X. Tata, Phys. Rev. D52,
2746 (1995).
\item[{23.}] M. Guchait and D.P. Roy, Phys. Rev. D52, 133 (1995).
\item[{24.}] See e.g. M. Carena, M. Olechowski, S. Pokorski and
C.E.M. Wagner, Nucl. Phys. B426, 269 (1994).
\item[{25.}] J. Bagger, K. Matchev, D. Pierce and R. Zhang,
Nucl. Phys. B491, 3 (1997).
\item[{26.}] J.L. Feng, K.T. Matchev and T. Moroi,
Phys. Rev. Lett. 84, 2322 (2000); Phys. Rev. D61, 075005 (2000).
\item[{27.}] Utpal Chattopadhyay, Amitava Datta, Anindya Datta,
Aseshkrishna Datta and D.P. Roy, Phys. Lett. B493, 127 (2000).
\item[{28.}] U. Chattopadhyay, T. Ibrahim and D.P. Roy,
Phys. Rev. D64, 013004 (2001).
\item[{29.}] J.L. Feng, K.T. Matchev and F. Wilczek,
Phys. Rev. Lett. 84, 2322 (2000); Phys. Rev. D63, 045024 (2001).
\item[{30.}] U. Chattopadhyay (private communication).
\item[{31.}] See e.g. U. Chattopadhyay and Pran Nath [hep-ph/0208012],
J.R. Ellis, A. Ferstl and K. Olive, Phys. Lett. B532, 318 (2002);
H. Baer et al. [hep-ph/0210441].
\item[{32.}] H. Baer, C.H. Chen, M. Drees, F. Paige and X. Tata,
Phys. Rev. D59, 055014 (1999); I. Hinchliffe and F. Paige,
Phys. Rev. D61, 095011 (2000).
\item[{33.}] M. Guchait and D.P. Roy, Phys. Lett. B541, 356 (2002).
\end{enumerate}

\newpage

\hrule width 0pt
\vspace*{1.5in}
\begin{figure}[h]
\includegraphics{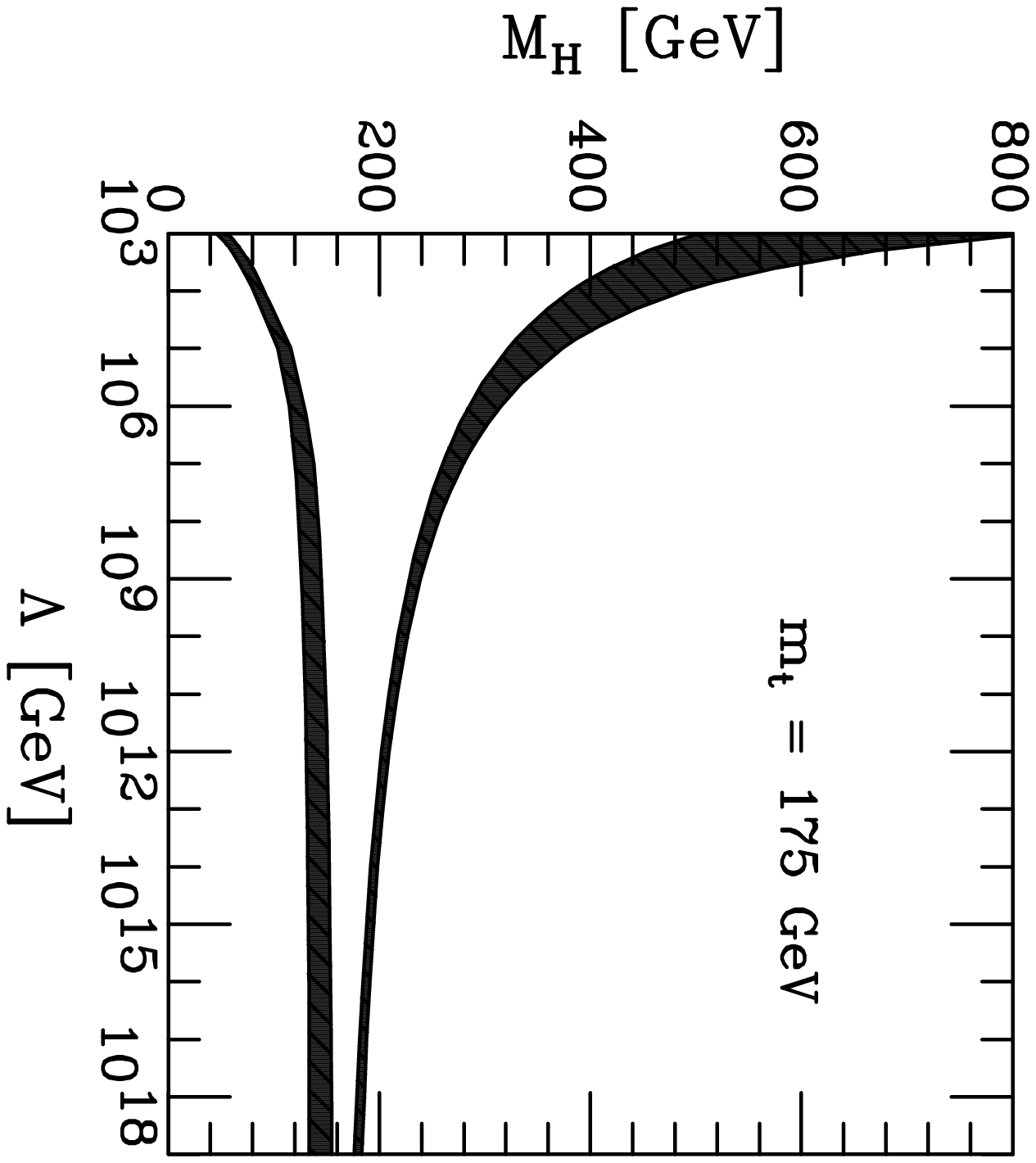}  
\label{fig:03-04fig1}
\end{figure}
\noindent Figure 1. The upper and lower bounds on the mass of the SM
Higgs boson as functions of the cutoff scale [4].

\vspace*{4.3in}
\begin{figure}[h]
\includegraphics{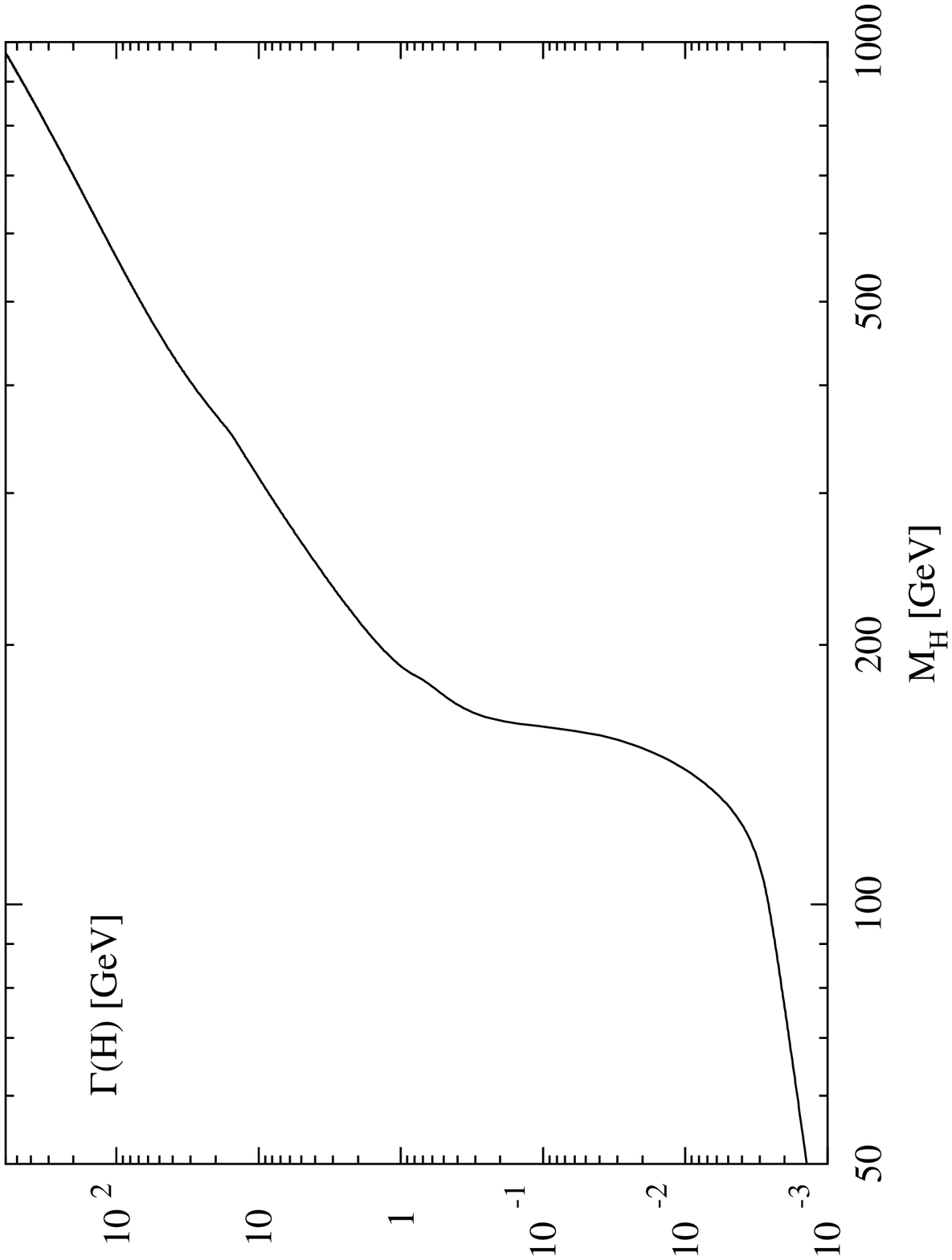}   
\label{fig:03-04fig2a}
\includegraphics{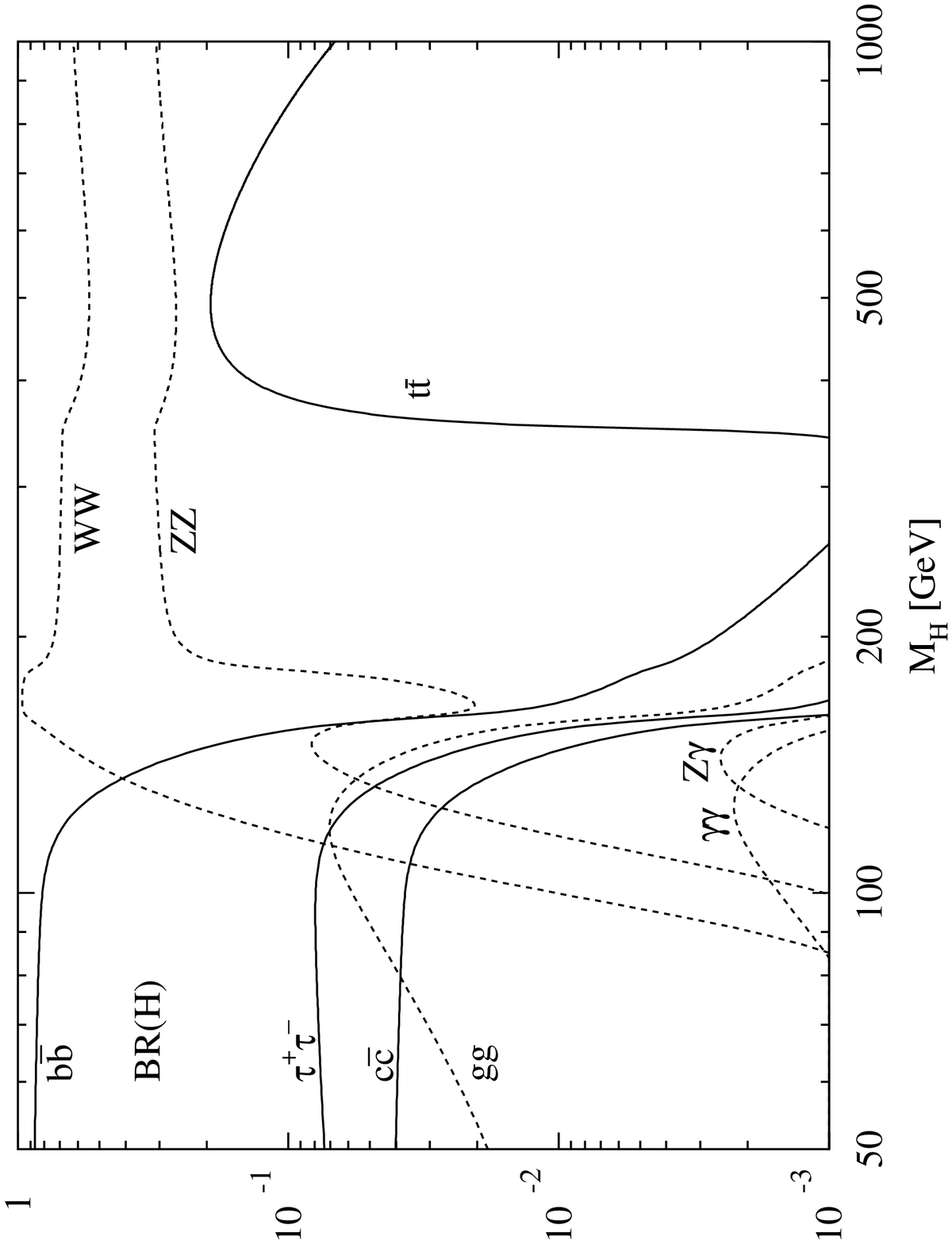}   
\label{fig:03-04fig2b}
\end{figure}
\noindent Figure 2. Total decay width and the main branching ratios of
the SM Higgs boson [8].

\newpage

\hrule width 0pt
\vspace*{2.5in}
\begin{figure}[h]
\includegraphics{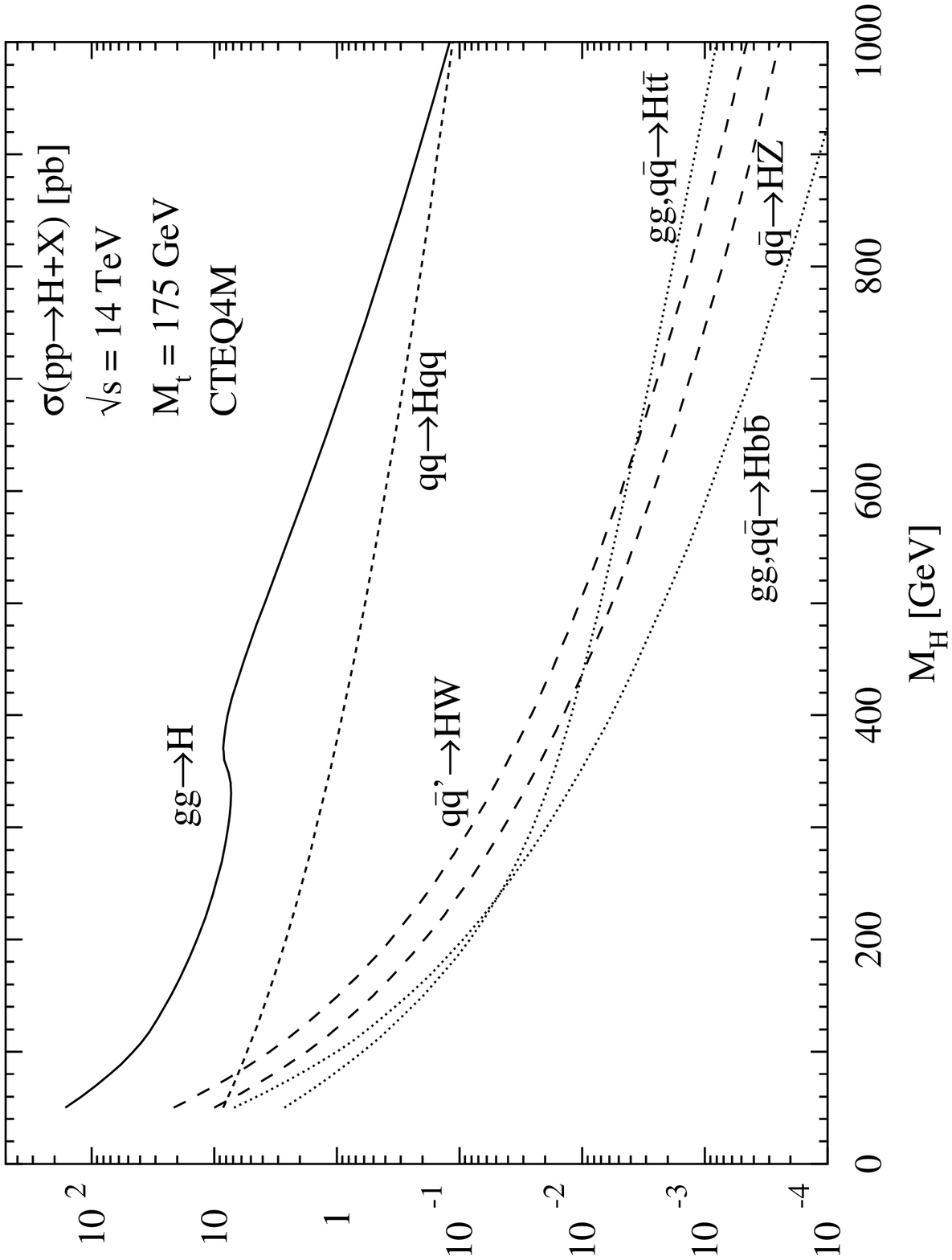}  
\label{fig:03-04fig3}
\end{figure}
\noindent Figure 3. Production cross-sections of the SM Higgs boson at
LHC [8].

\vfill

\begin{figure}[h]
\includegraphics{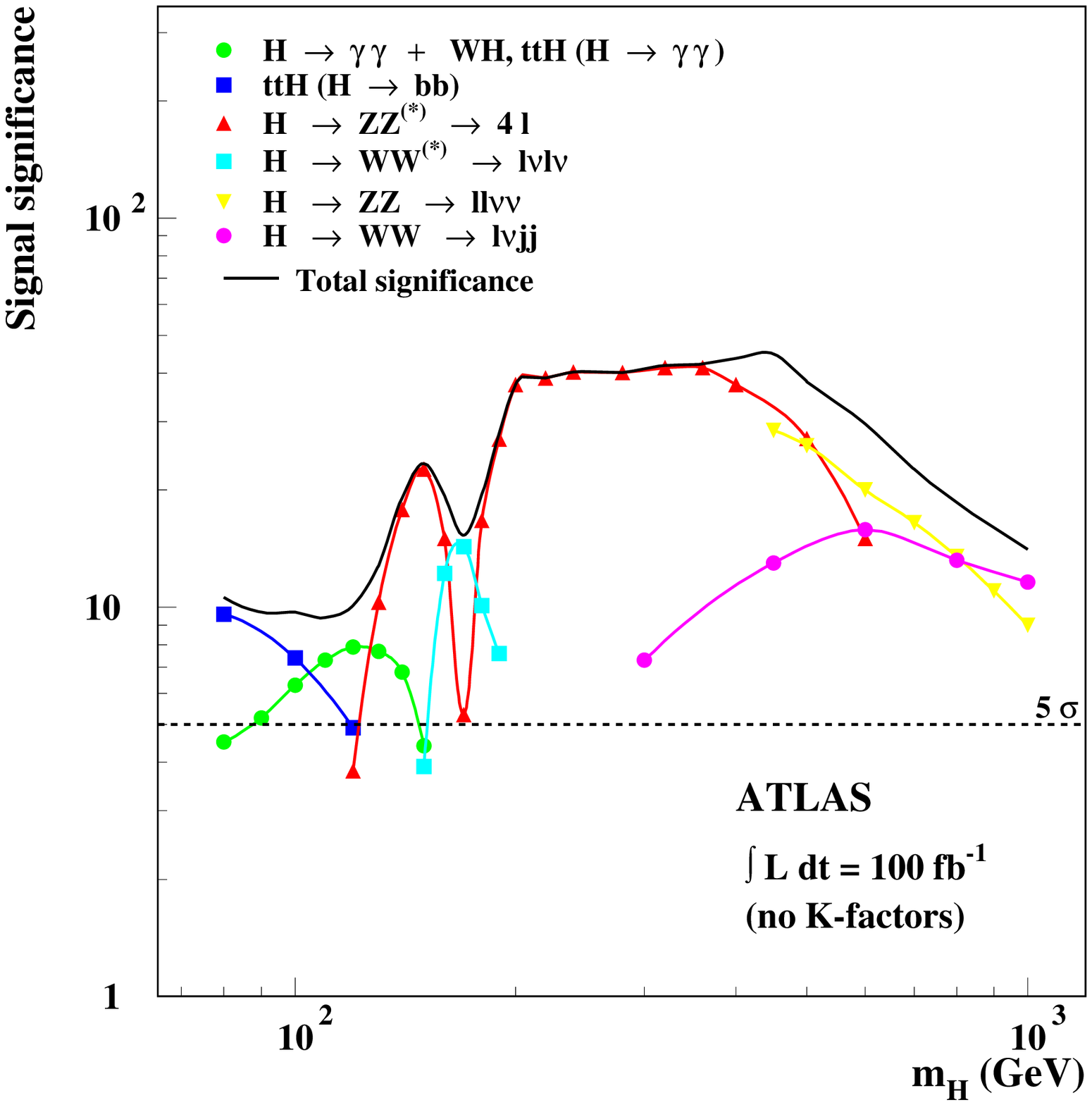}  
\label{fig:03-04fig4a}
\includegraphics{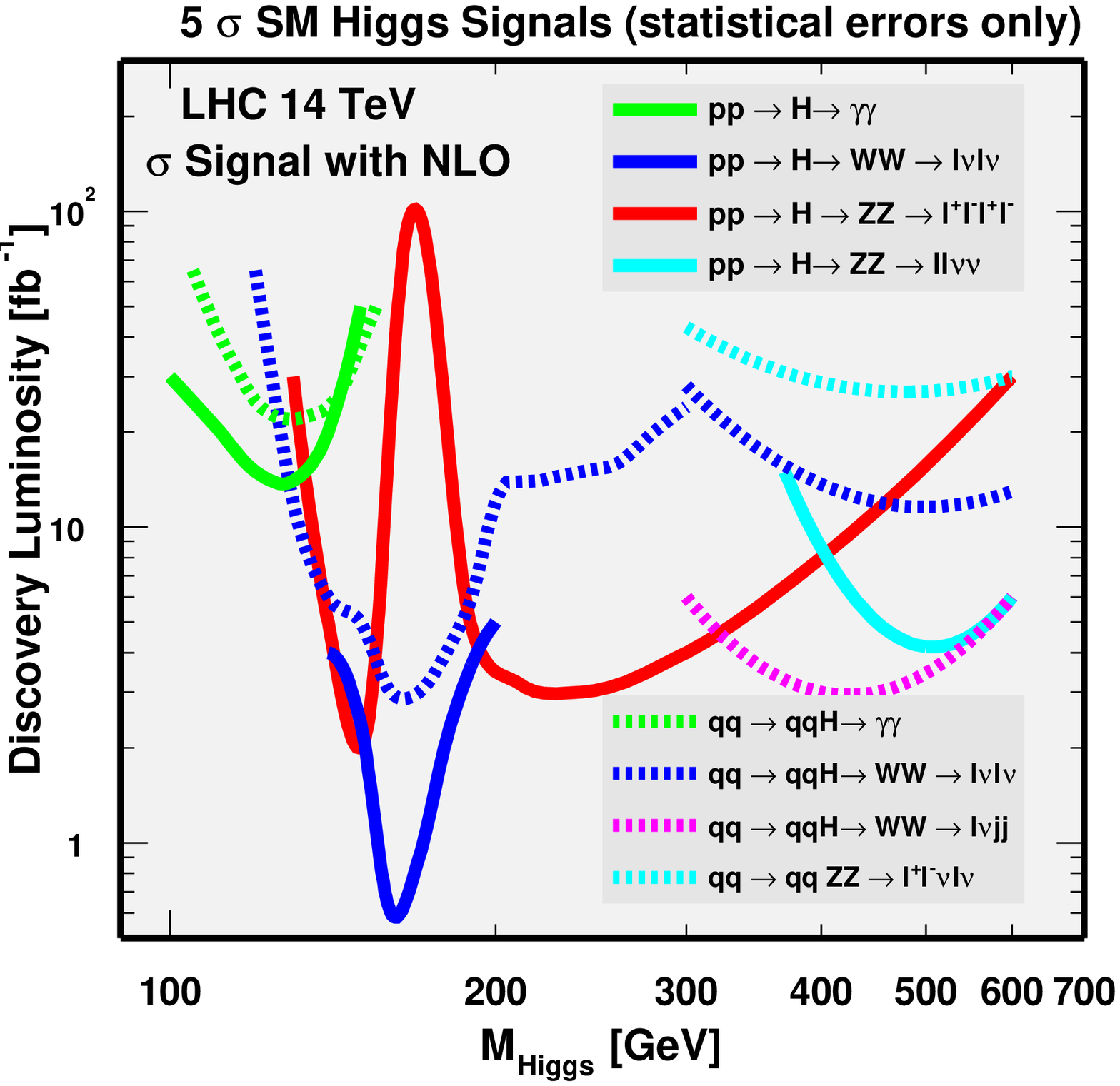}  
\label{fig:03-04fig4b}
\end{figure}
\noindent Fig. 4. (a) Significance level of the SM Higgs signal at LHC
with a Luminosity of 100 fb$^{-1}$; (b) Required luminosity for a
$5\sigma$ Higgs signal at LHC.  The first and the second figures are
from the simulations of the ATLAS and CMS collaborations respectively [9].

\newpage

\hrule width 0pt
\vspace*{2in}
\begin{figure}[h]
\includegraphics{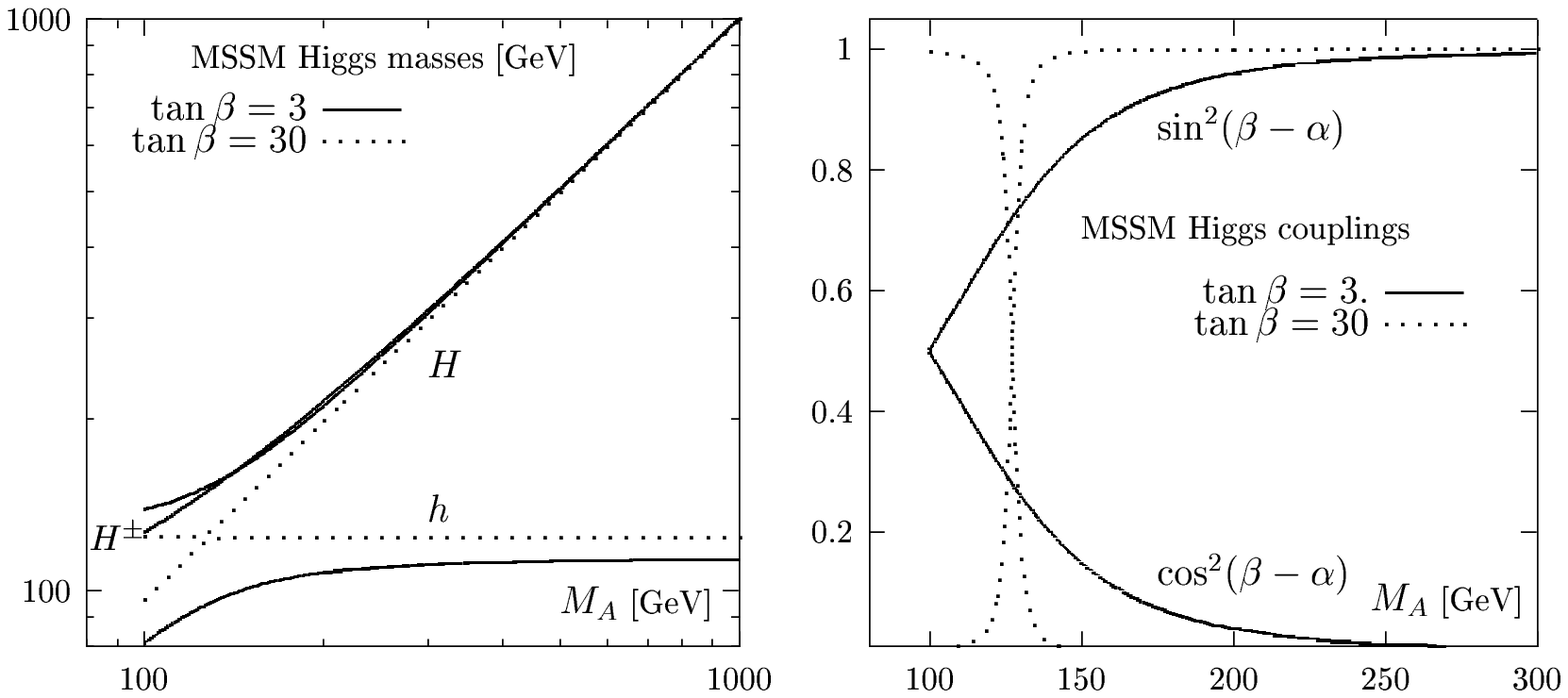}  
\label{fig:03-04fig5}
\end{figure}
\noindent Fig. 5. Masses of the MSSM Higgs bosons and their squared
couplings to $WW$, $ZZ$ (relative to the SM Higgs coupling) for two
representative values of $\tan\beta = 3$ and 30, assuming maximal stop
mixing [13].

\vfill

\begin{figure}[h]
\includegraphics{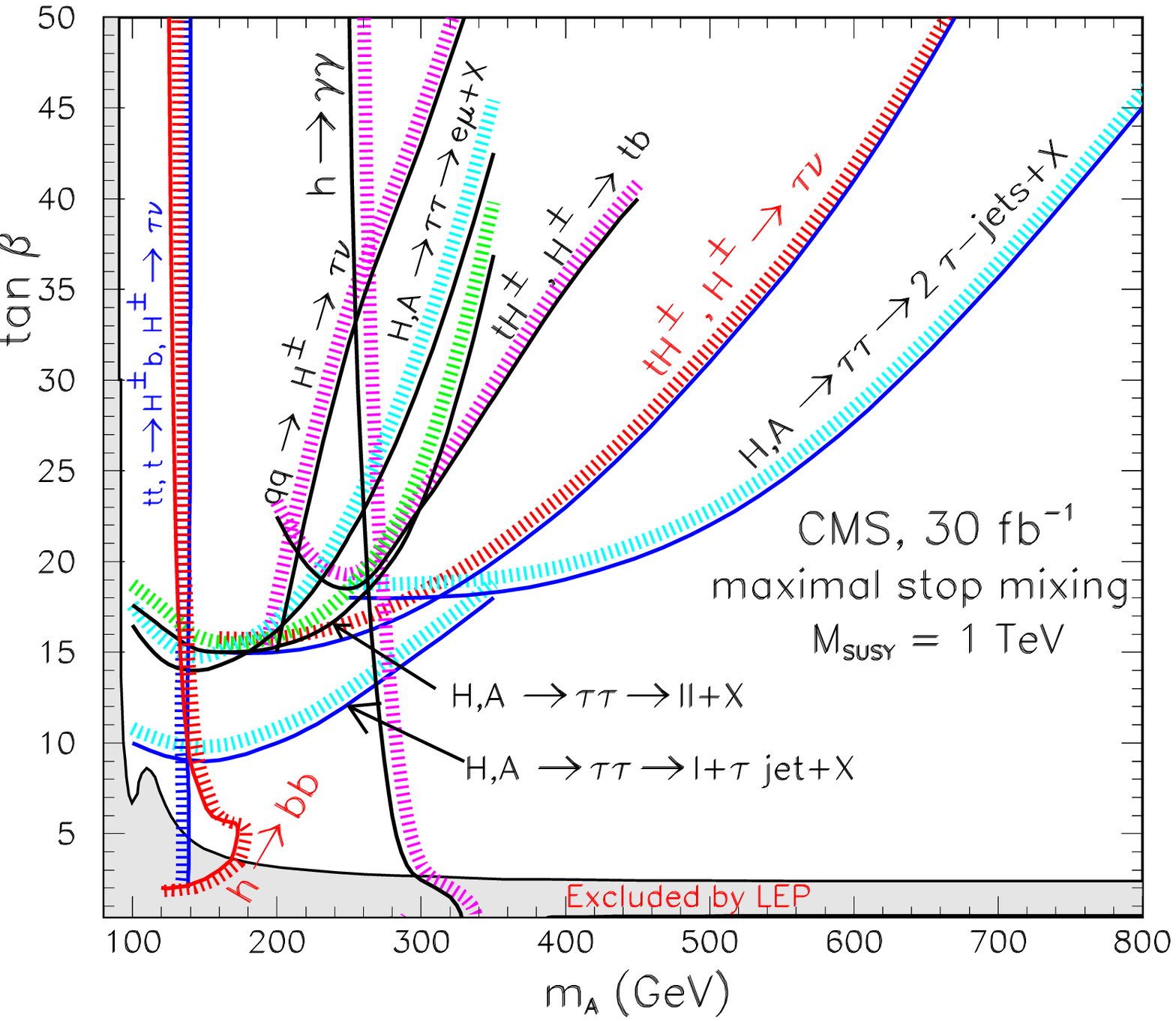}  
\label{fig:03-04fig6a}
\includegraphics{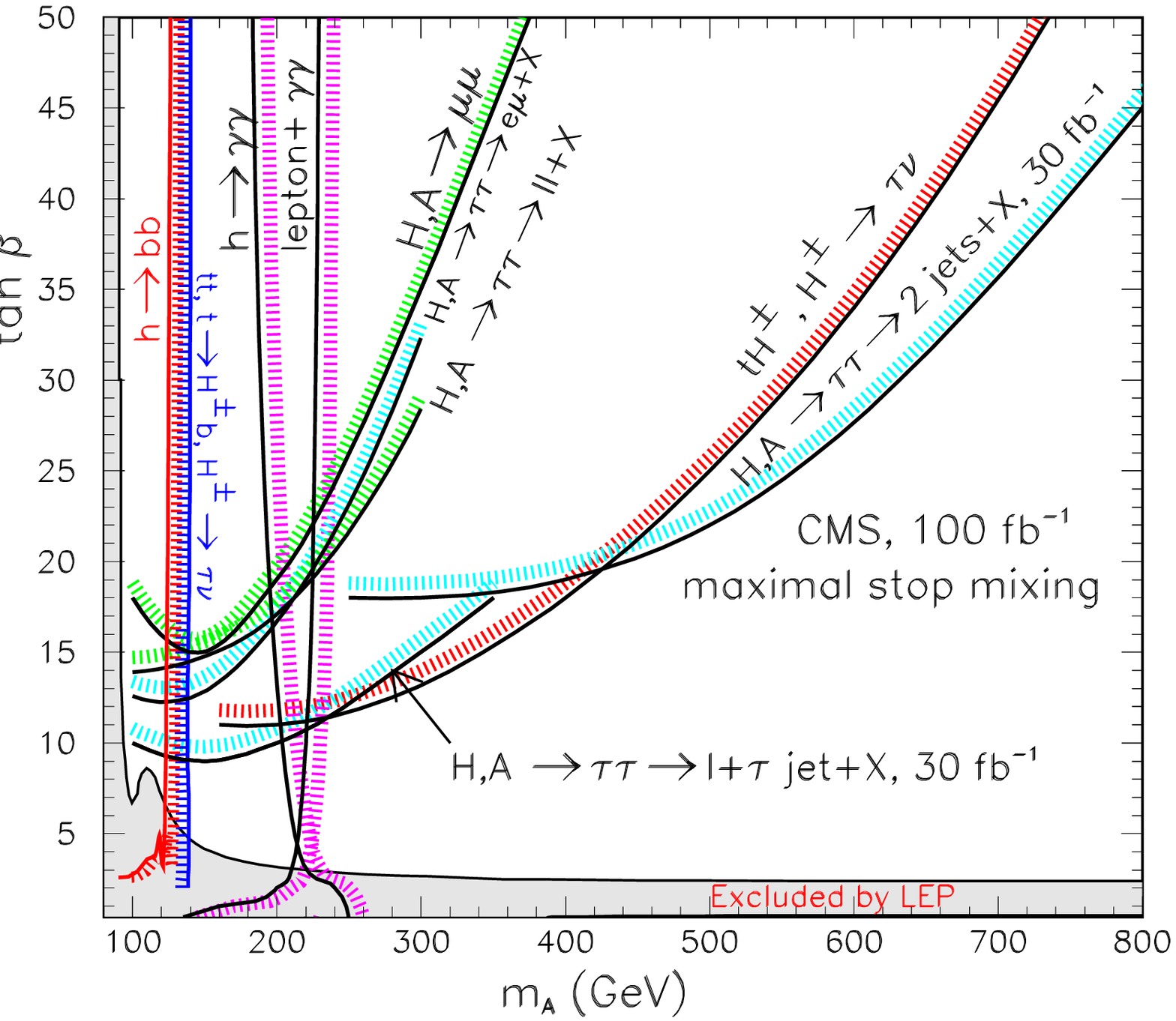}  
\label{fig:03-04fig6b}
\end{figure}

\noindent Fig. 6. Expected $5\sigma$ discovery limits of various MSSM
Higgs signals at LHC for luminosities of 30 fb$^{-1}$ and 100
fb$^{-1}$ [14].

\newpage

\begin{figure}[h]
\includegraphics{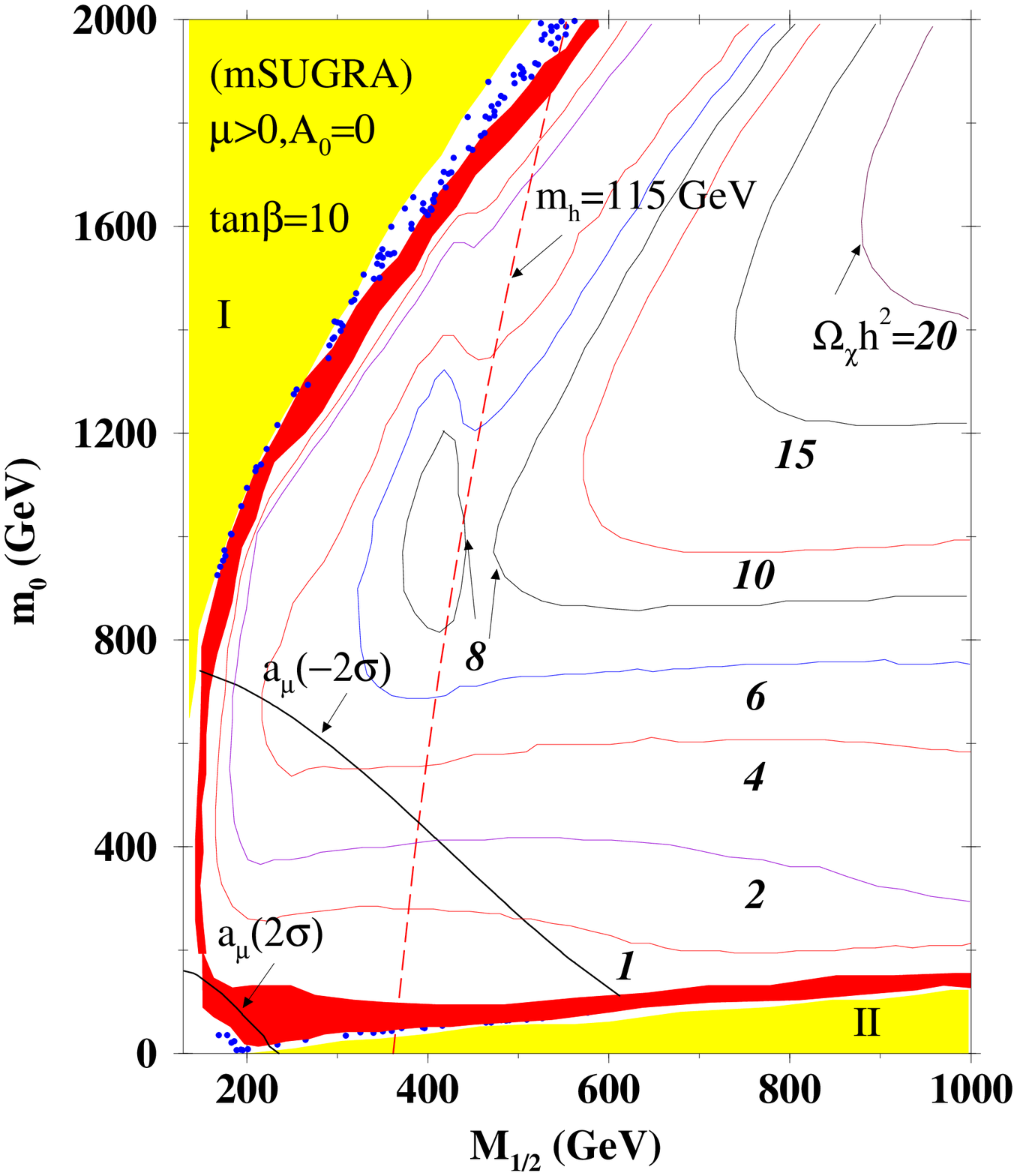}  
\label{fig:03-04fig7}
\end{figure}

\vfill

\noindent Fig. 7. Contours of DM relic density in the $m_0 - M_{1/2}$
plane for $\tan\beta = 10$ and +ve $\mu$ [30].  The excluded regions I
and II correspond to $\mu < 100$ GeV and $m_{\tilde\tau_1} <
m_{\tilde\chi^0_1}$ respectively.  Also shown are the exclusion limits
from the putative muon anomalous magnetic moment.

\newpage

\begin{figure}[h]
\includegraphics{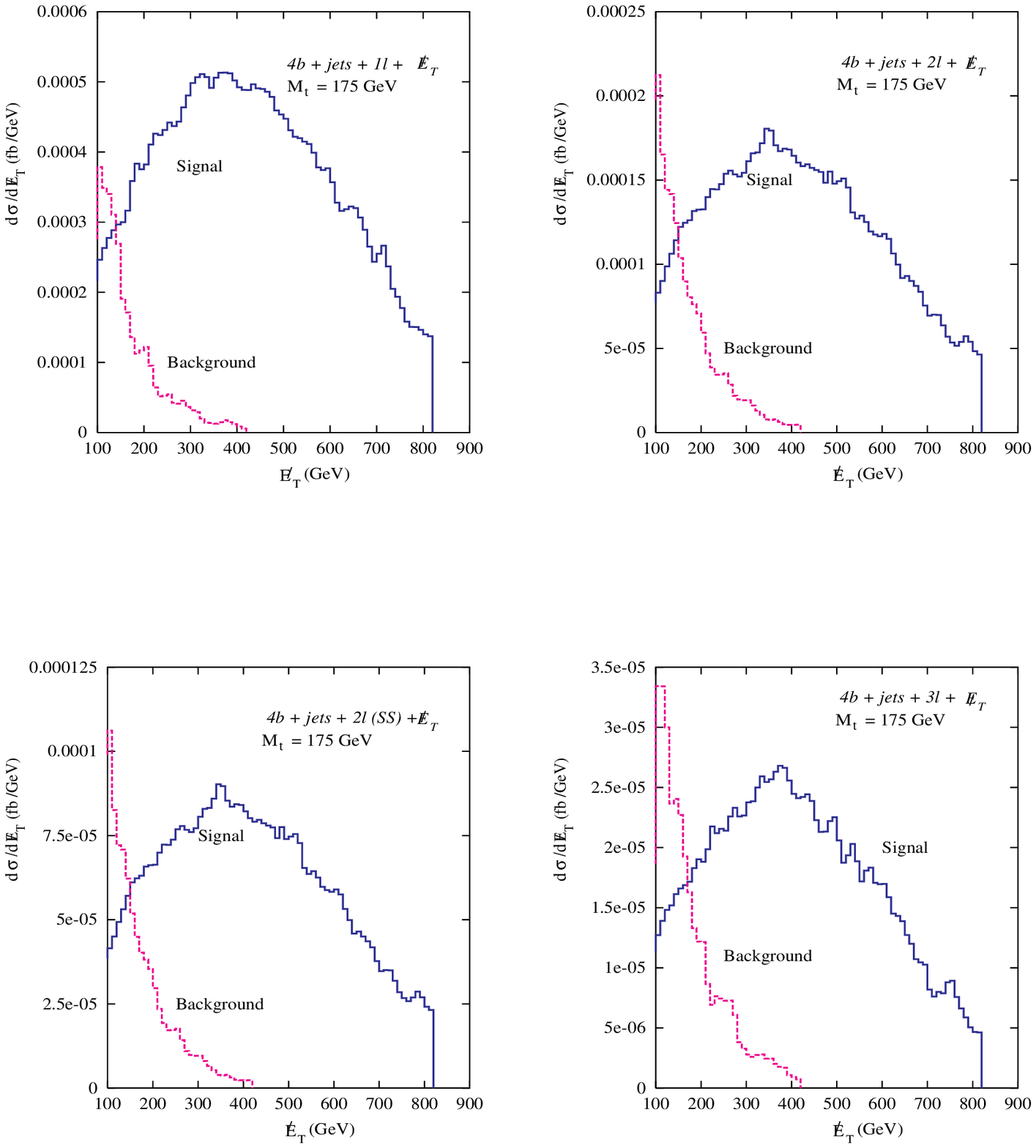}  
\label{fig:03-04fig8}
\end{figure}

\vfill

\noindent Fig. 8. Expected SUSY signal and the irreducible SM
background from $t\bar t$ $t\bar t$ are shown in the single lepton,
dilepton, same sign dilepton and trilepton channels with $\geq 3$
$b$-tags for the focus point region ($m_0 = 2$ TeV, $M_{1/2} = 500$
GeV, $\tan\beta = 10$) [27].

\newpage

\begin{figure}[h]
\includegraphics{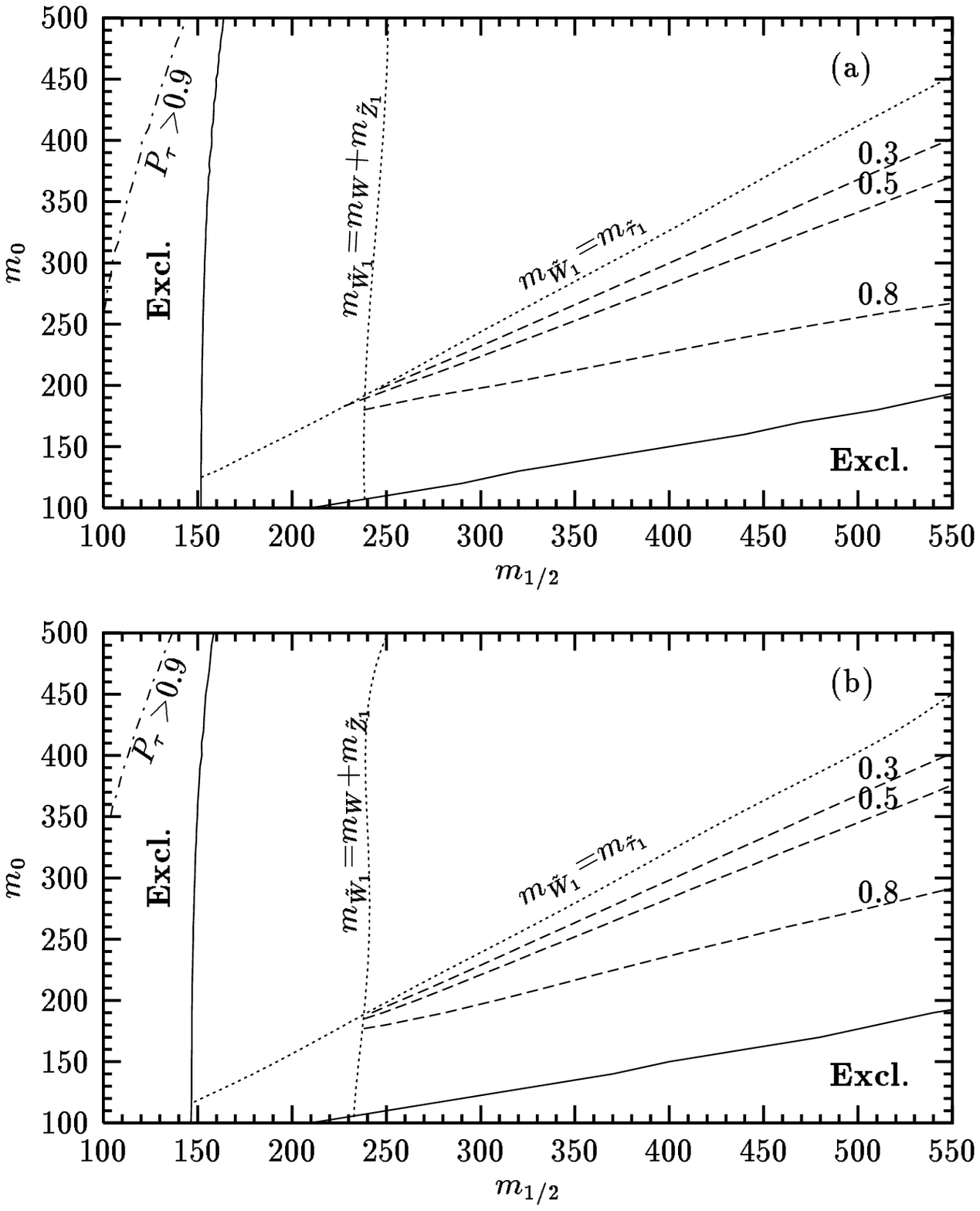}  
\label{fig:03-04fig9}
\end{figure}

\vfill

\noindent Fig. 9. $BR(\tilde\chi^\pm_1 \rightarrow \tilde\tau_1
\rightarrow \tau \chi^0_1)$ is shown in the $m_0 -M_{1/2}$ plane for
$A_0 = 0$, $\tan\beta = 30$ and (a) positive $\mu$, (b) negative
$\mu$.  The entire region to the right of the dot-dashed line
corresponds to $P_\tau > 0.9$ [33].

\newpage

\begin{figure}[h]
\includegraphics{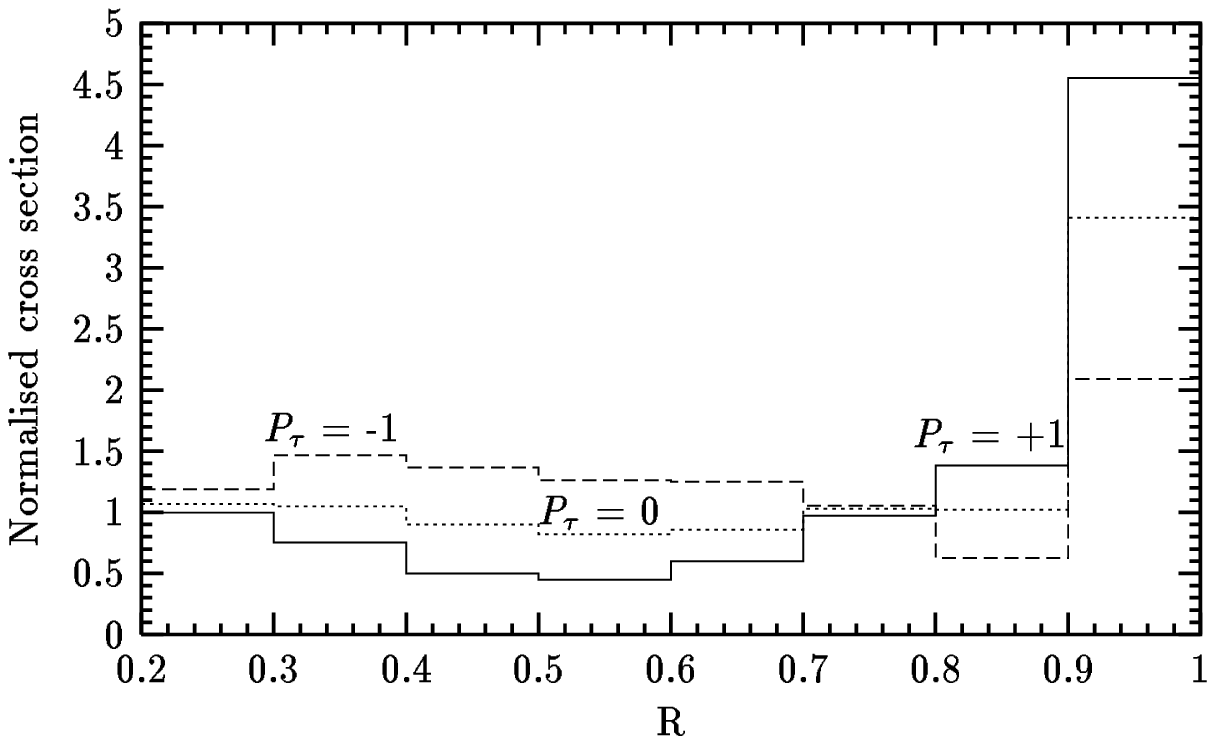}  
\label{fig:03-04fig10}
\end{figure}

\vspace*{3.5in}

\noindent Fig. 10. The normalised SUSY signal cross-sections are shown
for $P_\tau = 1$ (solid), 0 (dotted) and -1 (dashed) in the 1-prong
hadronic $\tau$-jet channel as functions of the $\tau$-jet momentum
fraction (R) carried by the charged prong [33].

\end{document}